\documentclass[12pt]{article}
\usepackage[dvips]{epsfig}
\usepackage{cite,amsmath,amssymb,dcolumn}
\newcommand{\mul}[1]{\multicolumn{1}{c}{#1}}
\begin{document}
\title{Multi-scale modeling of poly(isoprene) melts}
\author{Roland Faller\footnotemark \and Florian M{\"u}ller-Plathe\\
  \small Max-Planck-Institut f\"ur Polymerforschung, Ackermannweg 10,
  55128 Mainz, Germany}
\maketitle
\setcounter{footnote}{1}
\renewcommand{\thefootnote}{\fnsymbol{footnote}}
\abstract{
  \noindent Atomistic (atom-scale) and coarse-grained (meso-scale)
  simulations of structure and dynamics of poly-isoprene melts are
  compared. The local structure and chain packing is mainly determined
  by the atomistic details of the polymer architecture. The large-time
  dynamics encountered in NMR experiments can be explained by
  meso-scale simulations including stiffness. The connecting link
  between the two scales is the stiffness which, although being a local 
  property,
  influences strongly even the long-timescale dynamics. The standard reptation
  scenario fails to explain the observed dynamics. We
  propose {\it strong reptation} as a modified reptation scenario
  in which the local Rouse motion is absent. }
\footnotetext{present address: Department of Chemical Engineering, University 
  of Wisconsin-Madison, 1415 Engineering Drive, Madison, WI 53706}
\vspace{1ex}

\section{Introduction}
The large abundance of polymers and their variety of applications make them an
interesting target of study in theoretical material science. The understanding
of the differences and similarities of the various materials is an important
prerequisite for the goal of CAMD - computer aided materials design.
The dream of chemical engineers would be to develop in the workstation the 
perfect material for a given purpose.

To come closer to this distant goal, much effort in various
fields is necessary.  As polymeric materials are characterized by the
importance of various length scales, the understanding of the interplay
between these scales is of utmost importance. Methods adapted
to all relevant scales are needed from the experimental as well as from
the theoretical or simulational viewpoint. In simulations, much work has
been directed recently of the issue of {\it
coarse-graining}, the mapping of simulations on different scales, in order to
get a unified view of the arising
scales~\cite{tschoep98a,tschoep98b,baschnagel00,meyer00,reith00s}. In
the present contribution we show that atomistic simulations of {\it
trans}-polyisoprene (PI)~\cite{faller00sb} can be mapped onto a simple
bead-spring model incorporating exclude volume, connectivity, and
stiffness~\cite{faller99b,faller00b}. With this simple model we can
investigate the long-time dynamics of chains in the melt. There is
evidence for reptation which qualitatively changes with
stiffness~\cite{faller00a,faller00sa}. This connects to results of
modern NMR experiments~\cite{graf98} which we can reproduce
to a satisfactory extent~\cite{faller00b,mplathe00}. Thus, simulations on both
scales and especially their connection reveal different important aspects of 
the system under study.

The remainder of this contribution is organized as follows. In
section~\ref{sec:atomist} a short review about our recent results of
atomistic simulations of {\it Trans}-PI is given. In
section~\ref{sec:meso} results of simulations on the meso-scale level,
where the polymer identity is put into a simple stiffness parameter,
are presented. In the concluding section the important concept of
stiffness is discussed and it is shown that stiffness is sufficient to
allow a mapping for the polymer presented here.
\section{Local Structure and Dynamics - The atomistic scale}
\label{sec:atomist}
\begin{figure}
  \[
  \includegraphics[width=0.6\linewidth]{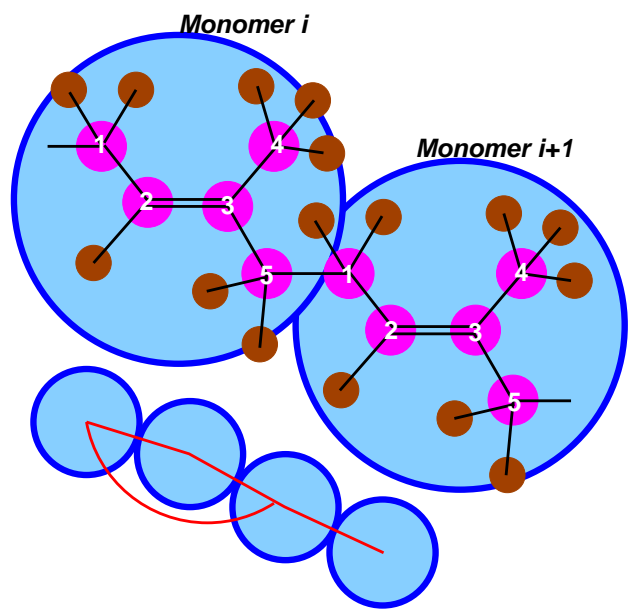}
  \]
  \caption{The atomistic structure of {\it Trans}-1,4-poly-isoprene and its
    mapping to a bead-spring model including only excluded volume and
    stiffness.} 
  \label{fig:sketch}
\end{figure}

The atomistic structure of oligomers of {\it trans}-1,4-polyisoprene
(cf. Figure~\ref{fig:sketch}) was investigated. For details of the
simulations and the interaction potentials see
refs.~\citen{faller00a,faller00sb}. Here we only note that the
simulation box contains 100 oligomers of average length 10 monomers of
pure {\it trans}-poly-isoprene which were pre-equilibrated using
end-bridging Monte Carlo~\cite{pant95,mavrantzas99} at the ambient
condition of 300~K or the elevated temperature of 413~K. All
atomistic simulations are run at 101.3~kPa.

\begin{figure}
  \includegraphics[angle=-90,width=0.49\linewidth]{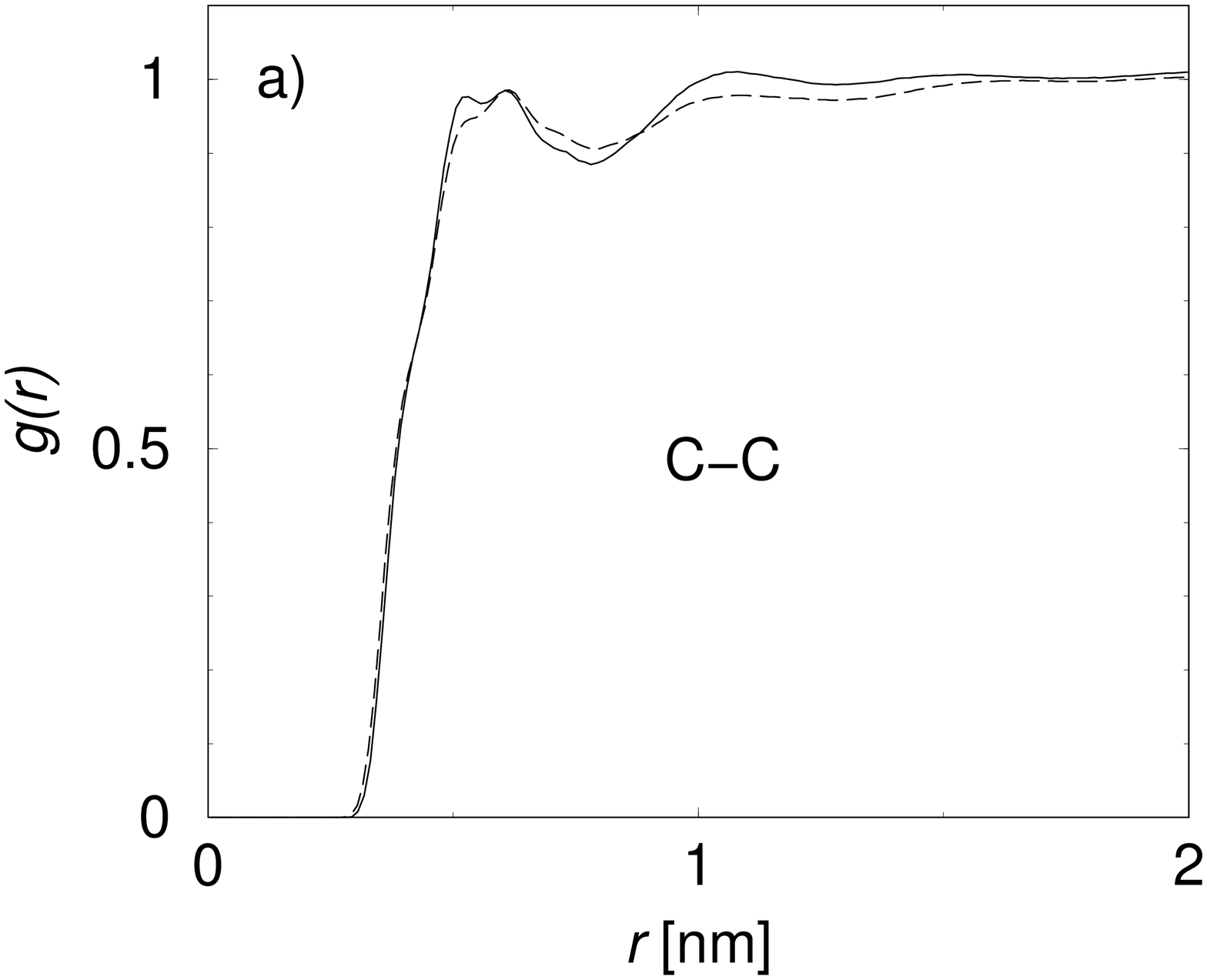}
  \includegraphics[angle=-90,width=0.49\linewidth]{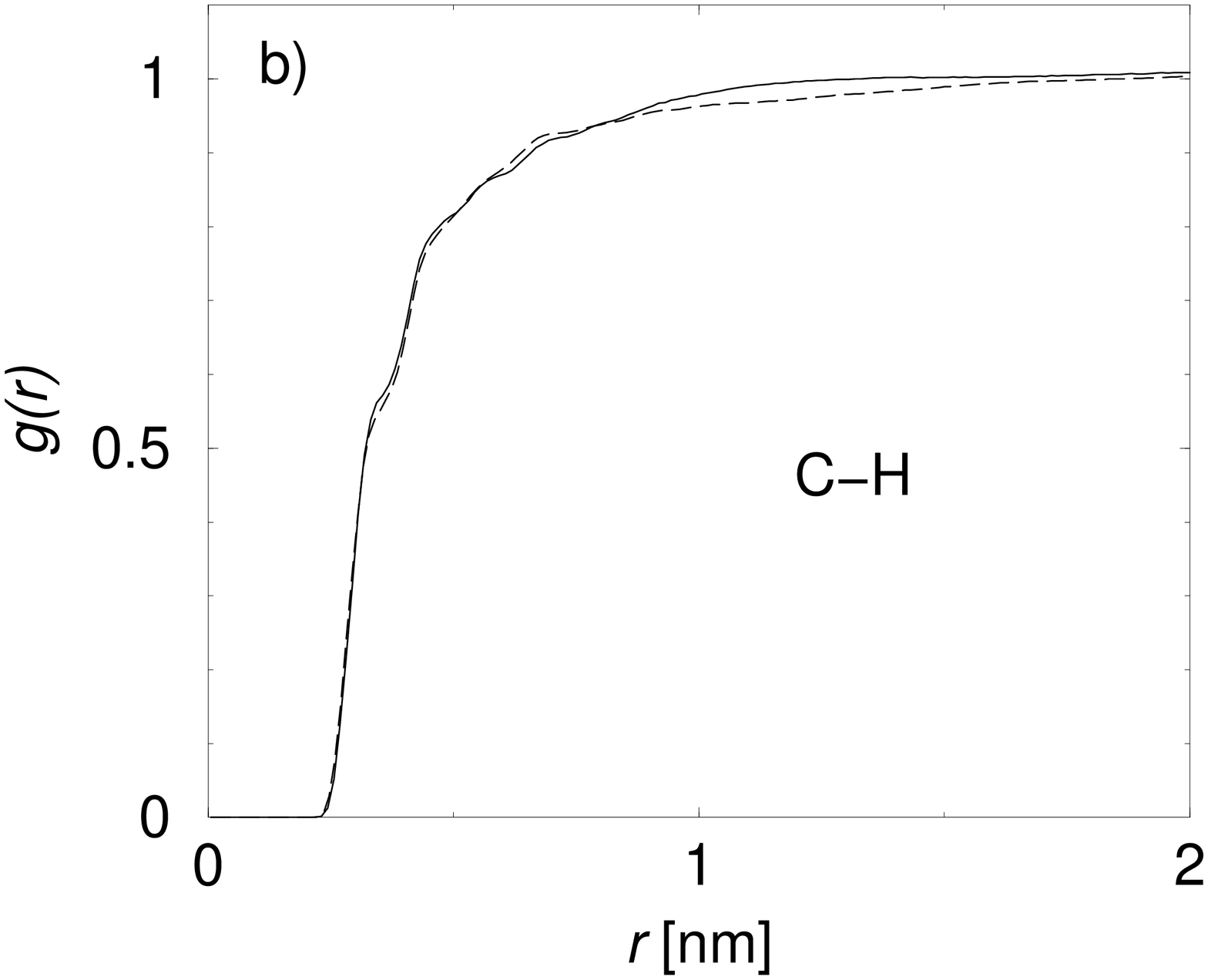}
  \includegraphics[angle=-90,width=0.49\linewidth]{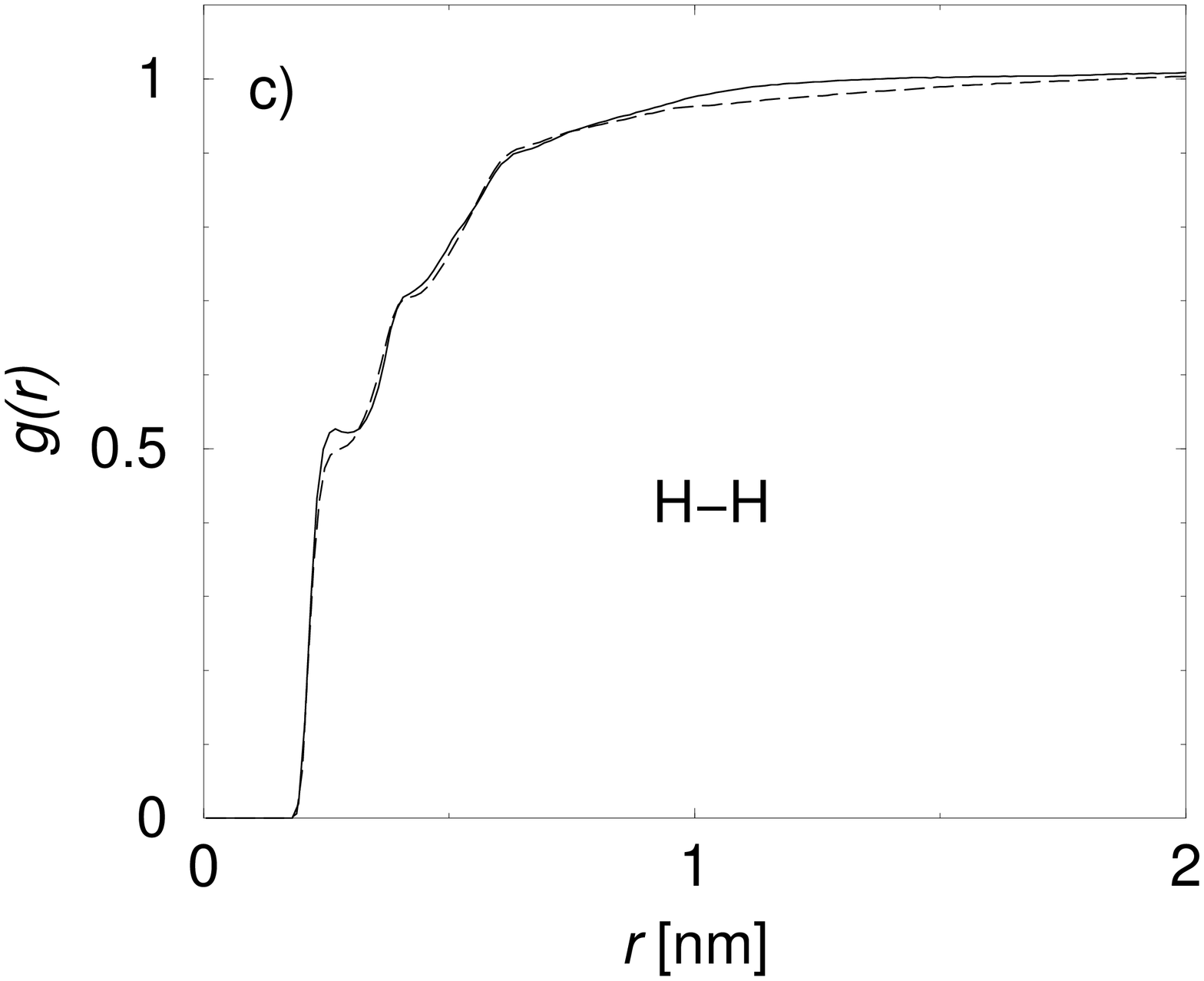}
  \caption{Radial distribution functions of atomistic data of
    {\it trans}-polysoprene. a) Carbon$-$Carbon, b) Carbon$-$Hydrogen and 
    c) Hydrogen$-$Hydrogen RDFs. The carbons show a stronger pronounced 
    structure. A correlation hole can be seen in subfigure c).}
  \label{fig:rdfato}
\end{figure}

The main result is, that the local structure is very important
for properties on the sub-monomer length
scale. Figure~\ref{fig:rdfato} shows the local mutual packing of
chains by means of radial distribution functions of the different
atoms in the melt. The carbons show a distinct peak at next neighbor
contact. The hydrogens, in contrast, show very little structure. Thus, we
encounter a carbon backbone with a surrounding ``hydrogen cloud''. This
can be taken as a first hint that not all details are necessary for
every simulation. Still, the hydrogens had to be taken into account
for the C$-$H vector reorientation. This is important for direct
comparison to NMR experiments (below). The comparison to experimental raw data
is always a good and necessary validation of simulation
models~\cite{mplathe00}.

\begin{figure}
  \[
  \includegraphics[angle=-90,width=0.5\linewidth]{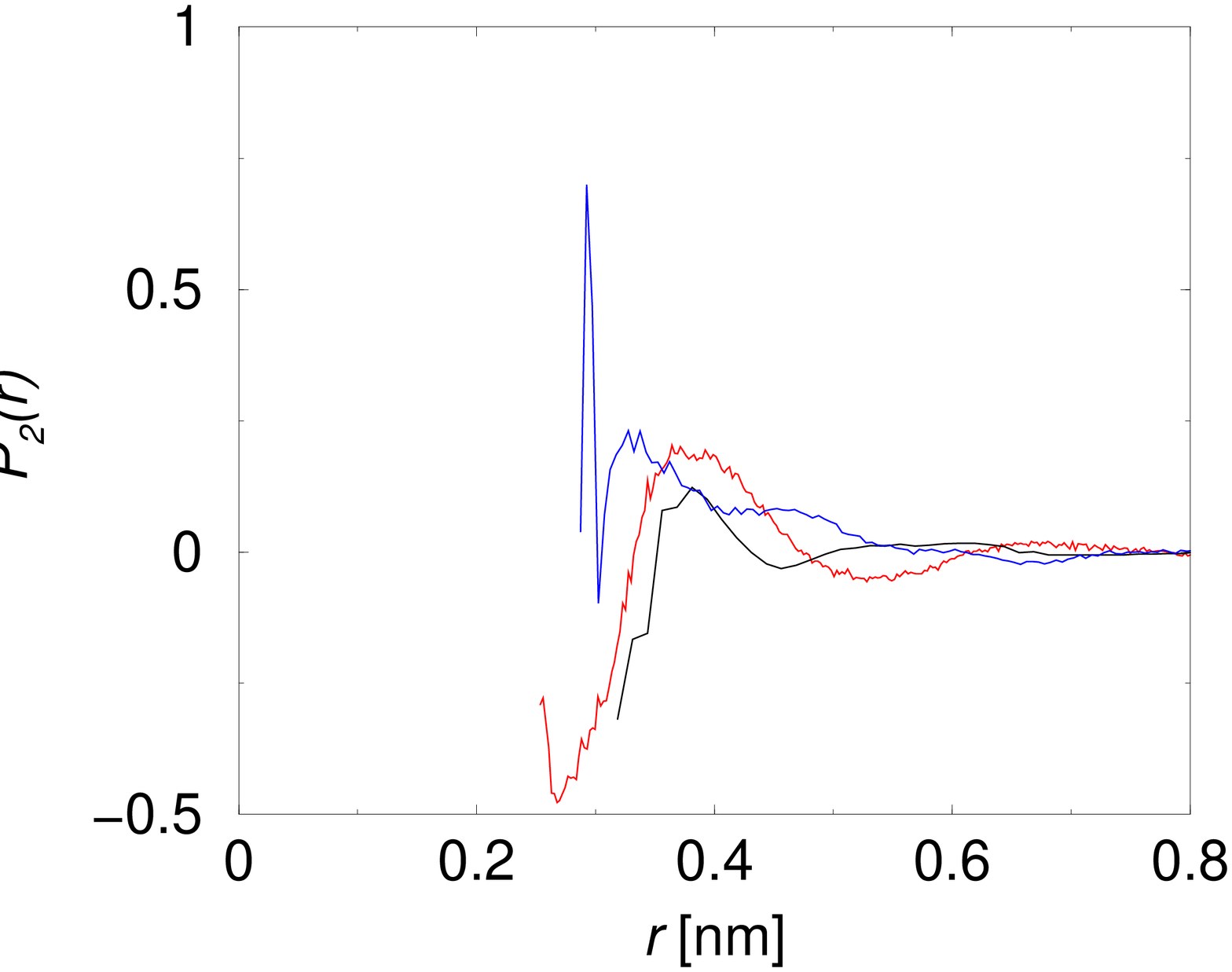}
  \]
  \caption{Atomistic mutual orientation of double-bonds (blue) in 
    trans-polyisoprene in
    comparison to the fully flexible simple model of section~\ref{sec:meso} 
    (red). The black line is for the atomistic vectors connecting C$_5$ with 
    C$_1$ of the next monomer.}
  \label{fig:atoodf}
\end{figure}

The local packing is additionally reflected by the directionality of
chains at contact. This is shown in Figure~\ref{fig:atoodf} by means of the
spatial orientation correlation function of the double bonds
\begin{equation}
  OCF(r):=P_2(r)=\Big\langle\frac{1}{2}
  \Big[3\cos^2\alpha(r)-1\Big]\Big\rangle, 
\end{equation}
where $\alpha(r)$ is the angle between tangent unit vectors on two different chains. The
distance $r$ is measured between their centers of mass. The unit vectors may
be defined in different manners, e.g. the double bonds in 
Figure~\ref{fig:atoodf} are one possibility to denote the direction of a
monomer.

Direct comparisons to experiments~\cite{batie89,denault89,zorn92} and
simulations~\cite{moe96a,moe99} on {\it cis}-polyisoprene proof that our model
is realistic~\cite{faller00sb,faller00a}. 

\begin{figure}
  \[
  \includegraphics[angle=-90,width=0.49\linewidth]{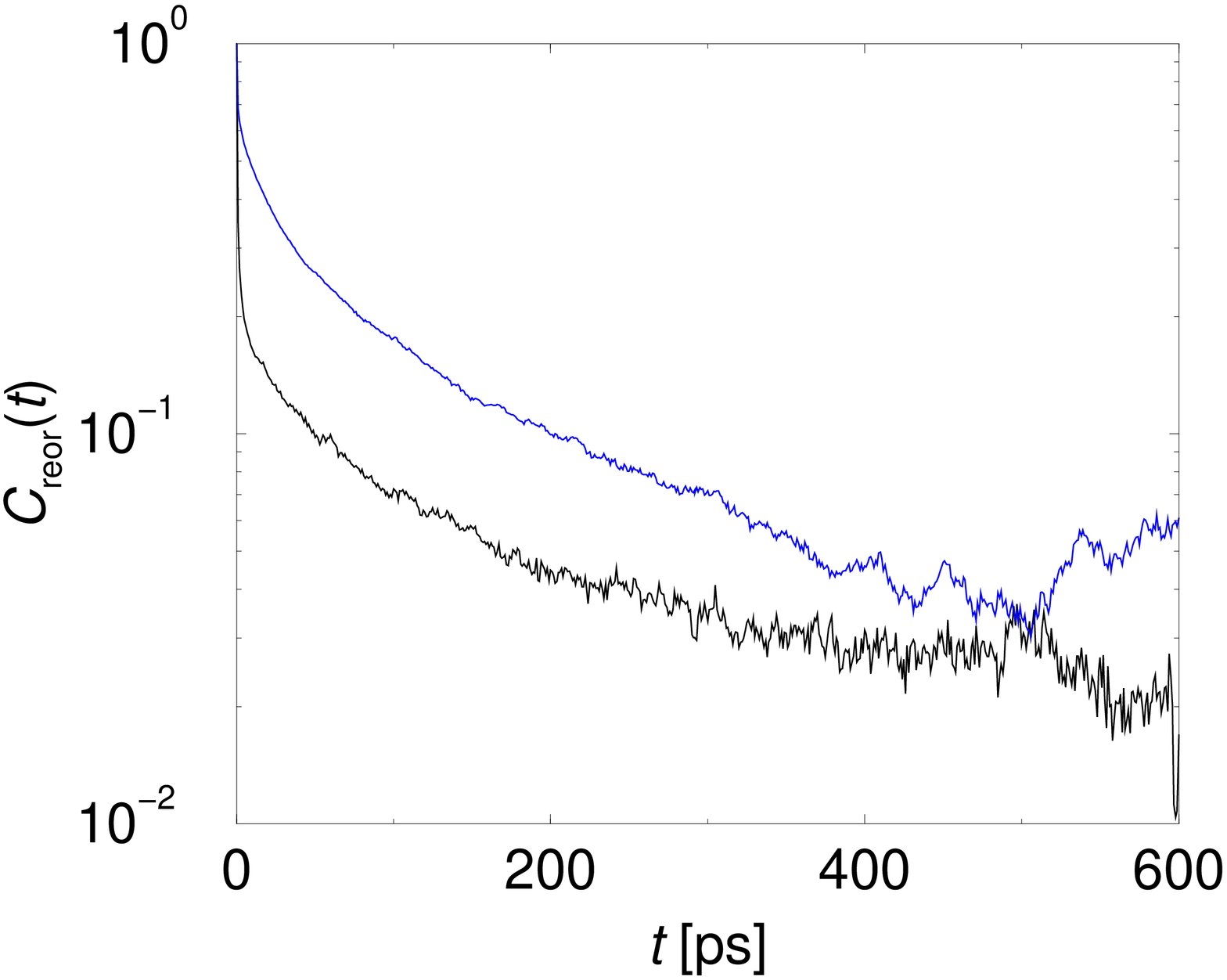}
  \]
  \caption{Reorientation correlation functions of the methylene C$-$H vectors
    at 413~K. Blue: C$_5-$H, Black: C$_1-$H.} 
  \label{fig:CHreor}
\end{figure}

The reorientation of the
hydrogens is subject to a  two-stage process (cf. Fig.~\ref{fig:CHreor}).
The first stage is a fast initial decay on the time-scale of the
segmental motion, i.e. the motion on the monomer or sub-monomer
scale. On the second time-scale the reorientation of the whole
oligomer becomes important. Overall, reorientation is monitored by the
following correlation function
\begin{equation}
  C_{\text{reor}}=\Big\langle 
  \frac{1}{2}\Big[3\cos^2\Phi(t)-1\Big]\Big\rangle.
  \label{eq:creor}
\end{equation}
Here, $\Phi(t)$ is the angle by which a given bond vector reoriented in
time $t$. The second Legendre polynomial is chosen as it is the
relevant quantity in NMR measurements. The effectivity of the two
stages may be measured by fitting double exponential decays to the
obtained correlation functions. In doing so, satisfactory ageement
with experiments could be achieved (cf.  Table~\ref{tab:avalues}) if
we keep in mind that the investigated systems are not completely
identical. The experiments were performed on mixtures with a high {\it
cis}-PI content and longer chains.
\begin{table}
  \[
  \begin{array}{c*{3}{D{.}{.}{-1}}}
    \hline
    \text{vector} 
    & \mul{a_{1\text{ps}}^{trans}} &  
    \mul{a_{\text{sim}}^{cis}}& \mul{a_{\text{exp}}^{cis}}\\ 
    \hline
    & \mul{\text{sim: this work}} 
    & \mul{\text{sim: ref. \citen{moe99}}} & 
      \mul{\text{exp: ref. \citen{batie89}}} \\
    \hline
    \operatorname{C}_{1}-\operatorname{H} & 0.42 & 0.28  & 0.40\\
    \operatorname{C}_{2}-\operatorname{H} & 0.16 & 0.16  & 0.17\\
    \operatorname{C}_{5}-\operatorname{H} & 0.18 & 0.23  & 0.48\\
    \hline
  \end{array}
  \]
  \caption{Comparison of the experimental ({\it cis}-PI) and simulation ({\it
      cis} and {\it trans}-PI) data for the efficiency of the
    initial stage of the reorientation process. The parameter $a$ denotes to
    which value the reorientation correlation function (Eq.~\ref{eq:creor})
    decays in the short-time process, i.e. before the long-term exponential 
    relaxation 
    sets in. $a_{1\text{ps}}$ is the value of $1-C_{\text{reor}}$ at 1ps.  
    In the analysis of the simulations for {\it cis}-polyisoprene a stretched
    exponential second process was assumed. The experiments used a range of
    temperature between 283~K and 363~K~\cite{batie89}. The {\it
     trans} simulations were at 300~K~\cite{faller00sb} and the 
    {\it cis} simulations at 363~K~\cite{moe99}.}
    \label{tab:avalues}
\end{table}

In section~\ref{sec:globdyn} comparisons of the long-time dynamics of the
corresponding simple model with experiments will be presented.
\section{Global Structure and Dynamics - Meso-scale simulations}
\label{sec:meso}
Simple polymer models allow one to investigate large-scale phenomena
both in time and space which are not accessible by atomistic
simulations. Therefore many researchers employed such models to look
for rather generic polymer properties or dynamical 
concepts~\cite{grest86,rigby88,kremer90,duenweg98,faller99d,puetz00}. One
of these important concepts is reptation~\cite{degennes71,doi86}. The
reptation concept explains much of the molecular weight dependence of
viscosity and the elastic and loss moduli~\cite{doi86}. However, this model
originates in the simple Rouse model~\cite{rouse53} where no stiffness
is included or it is subsumed into simple {\it Kuhn blobs}~\cite{kuhn34}.

In order to look for the influence of stiffness, a three-body potential for
stiffening the backbone has therefore to be
introduced~\cite{micka97,faller99b,faller00b} 
\begin{equation}
  \frac{V_{\text{angle}}}{k_BT}
	=\frac{l_p}{k_BT}\Big[1-\cos(\vec{u}_i\vec{u}_{i+1})\Big].
\end{equation}
The force constant $l_p$ in this choice of units has the same numerical value 
as the resulting intrinsic persistence length (below), and is therefore 
denoted as $l_p$.
\subsection{Statics}
\begin{figure}
  \[
  \includegraphics[angle=-90,width=0.5\linewidth]{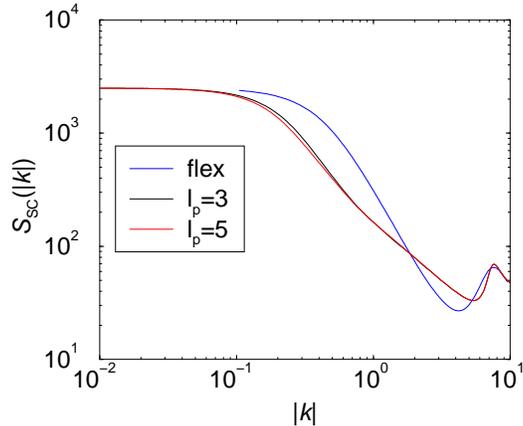}
  \]
  \caption{Static structure factor of model chains of length 50 monomers 
     depending on stiffness. Stiffness is given in terms of the persistence 
     length (see text). ``Flex'' means that no stiffening potential is 
     introduced; excluded volume leads to a persistence length of $l_p=1$.} 
  \label{fig:strfct}
\end{figure}

In polymer melts excluded volume is commonly assumed to be screened
out. Thus, we can expect that polymer chains behave as random
walks~\cite{doi86}. This is true at least on large length scales. All
{\it local} interactions only result in {\it local} chain
stretching. This is seen easily in the single chain static structure
factors of model chains~\cite{faller99b}
(cf. Figure~\ref{fig:strfct}). The random walk appears in the fractal
dimension of $d_f=2$ on the length scale bigger than the persistence
length. The fractal dimension expresses itself in the slope of the
structure factor. The chains with stiffness bend over to a weaker
slope indicating a smaller fractal dimension at higher $|k|$. The
persistence length $l_p$ originates from the Kratky-Porod worm-like
chain picture~\cite{kratky49}; $l_p$ measures the decay length of bond
correlations along the chain backbone.
Thus, static properties of different models and chains are easily
mapped onto each other by the simple {\it blob} concept where all local
interactions are put into one single length scale, which then allows a
renormalization onto flexible chains with the blobs acting as coarse-grained 
monomers. The size of the Kuhn blob $l_K$ and the persistence
length $l_p$ differ only by the constant factor of~2 although they
arise from different concepts.

\begin{figure}
  \[
  \includegraphics[angle=-90,width=0.5\linewidth]{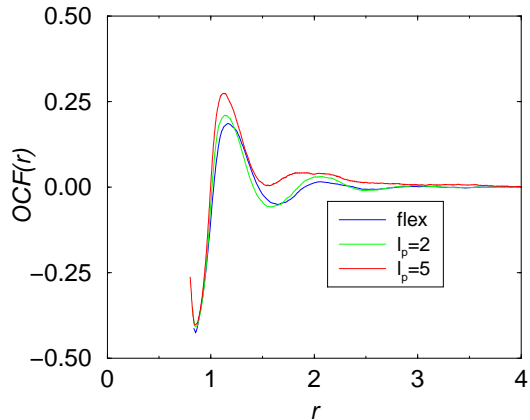}
  \]
  \caption{Static mutual chain packing of chains of different stiffness. The
    stiffer chains order more strongly parallel. Long range isotropy is
    preserved ($OCF\to0, r\to\infty$).} 
  \label{fig:odf}
\end{figure}

Still, even in chains of complete flexibility, i.e. no three-body
potential, there is a small but visible alignment between neighbouring
chains, similar to that shown in Figs.~\ref{fig:atoodf} and~\ref{fig:odf}
on the length
scale of about 3 chain diameters~\cite{faller99d}. This is evidence
that the very simple ansatz of model chains only incorporating
connectivity comes to its limits as soon as many-body interactions
come into play. As we have seen in section~\ref{sec:atomist} local
interactions affect strongly the mutual packing of chains. If chain
stiffness is increased, chain order becomes stronger~\cite{faller99b}
(cf. Fig.~\ref{fig:odf}) without leading to an overall nematic
order. This ordering is a strictly local phenomenon which is proven by
the fact that there is no chain length dependence
whatsoever~\cite{faller99d}.
\subsection{Dynamics}
\label{sec:globdyn}
\begin{figure}
  \[
  \includegraphics[angle=-90,width=0.5\linewidth]{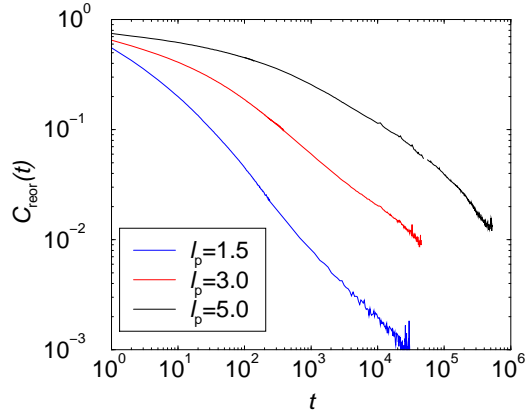}
  \]
  \caption{Algebraic reorientation of segments of the chain backbone depending
    on chain stiffness (Chain length: 200 monomers).}
  \label{fig:reoralg}
\end{figure}

\begin{figure}	
  \[
  \includegraphics[angle=-90,width=0.5\linewidth]{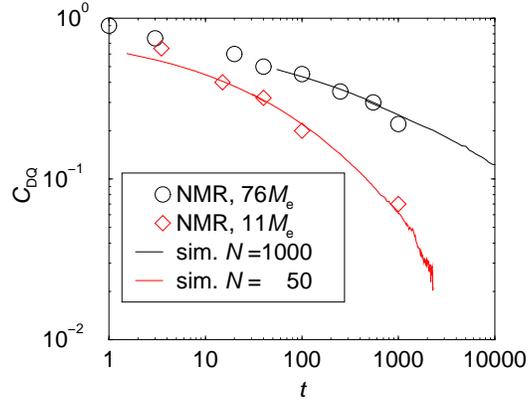}
  \]
  \caption{Comparison of simulated (chains of length 200 and $l_p=5$; lines) 
    and experimental (symbols)\cite{graf98} reorientation correlation 
    functions. For the experiments time-temperature superposition is assumed.}
\label{fig:cmpnmr}
\end{figure}

\begin{figure}
  \[
  \includegraphics[angle=-90,width=0.5\linewidth]{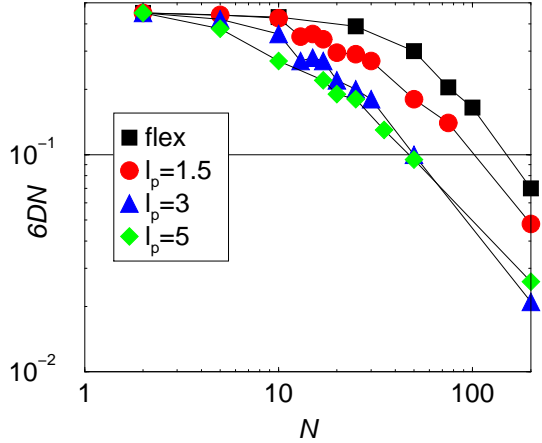}
  \]
  \caption{Diffusion coefficient of chains depending on chain length
    and stiffness. In the given representation Rouse behavior would
    give a straight horizontal line.}
  \label{fig:ne}
\end{figure}

\begin{figure}
  \[
  \includegraphics[angle=-90,width=0.5\linewidth]{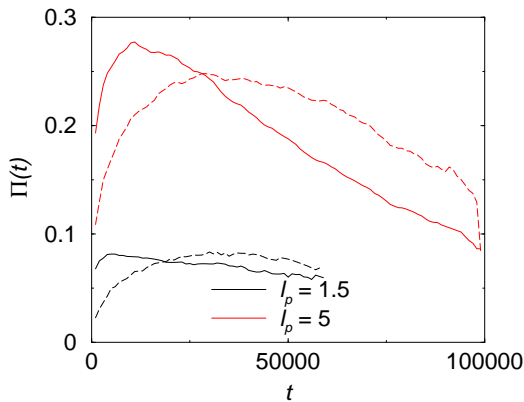}
  \]
  \caption{Correlations of monomer motion with the direction of the backbone
    depending on stiffness. Solid lines: $t=t_0$, dashed lines $t=t_0+\tau/2$} 
  \label{fig:anisomot}
\end{figure}

\begin{figure}
  \[
  \includegraphics[angle=-90,width=0.5\linewidth]{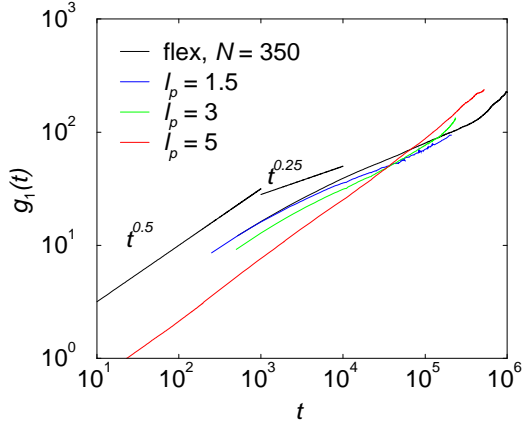}
  \]
  \caption{Mean squared displacements of central monomers in a chain depending
    on chain stiffness. For clarity lines indicating the slopes corresponding 
    to $t^{0.25}$ and $t^{0.5}$ are shown additionally.}
  \label{fig:msdcmp}
\end{figure}

\begin{figure}
  \includegraphics[angle=-90,width=0.5\linewidth]{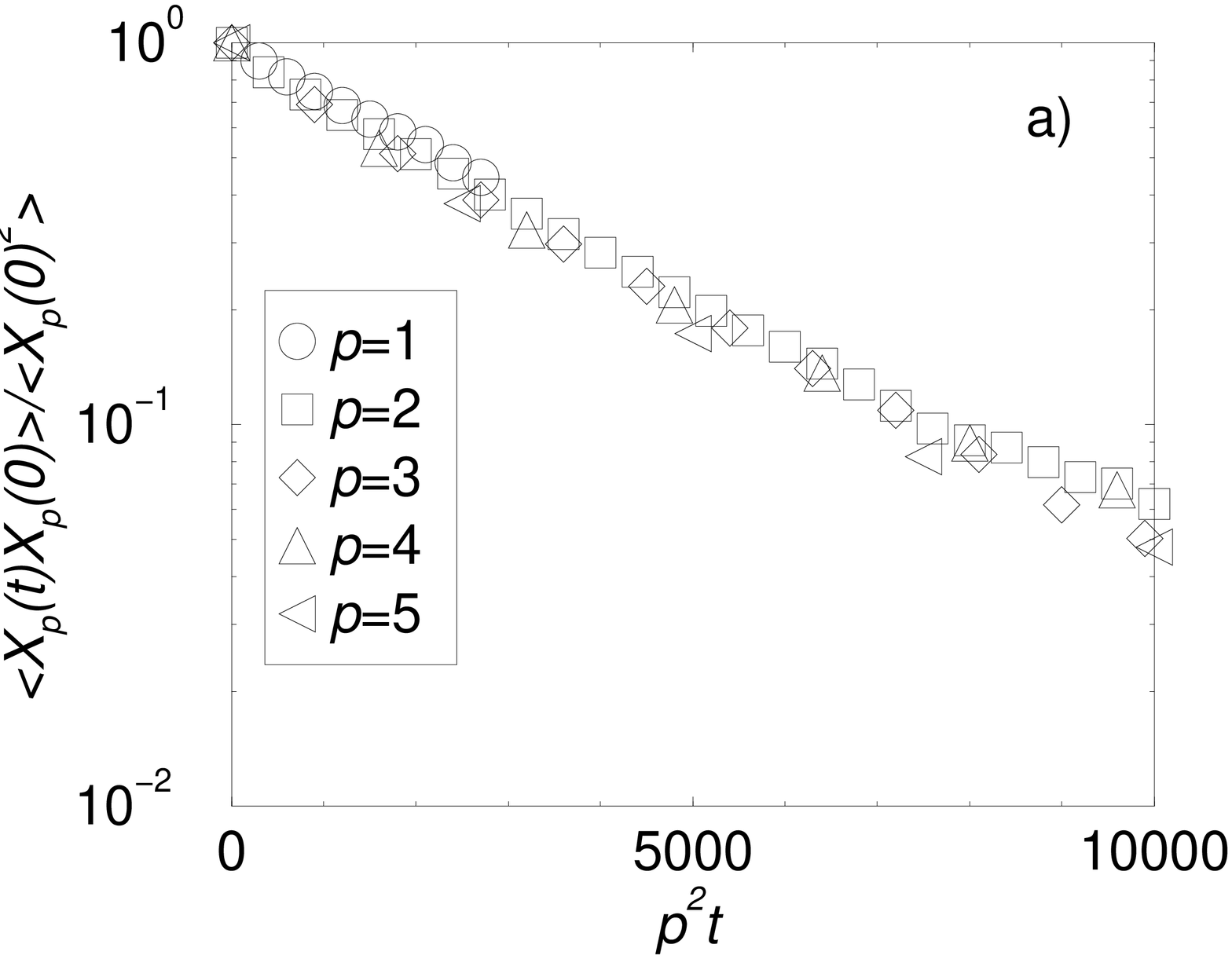}
  \includegraphics[angle=-90,width=0.5\linewidth]{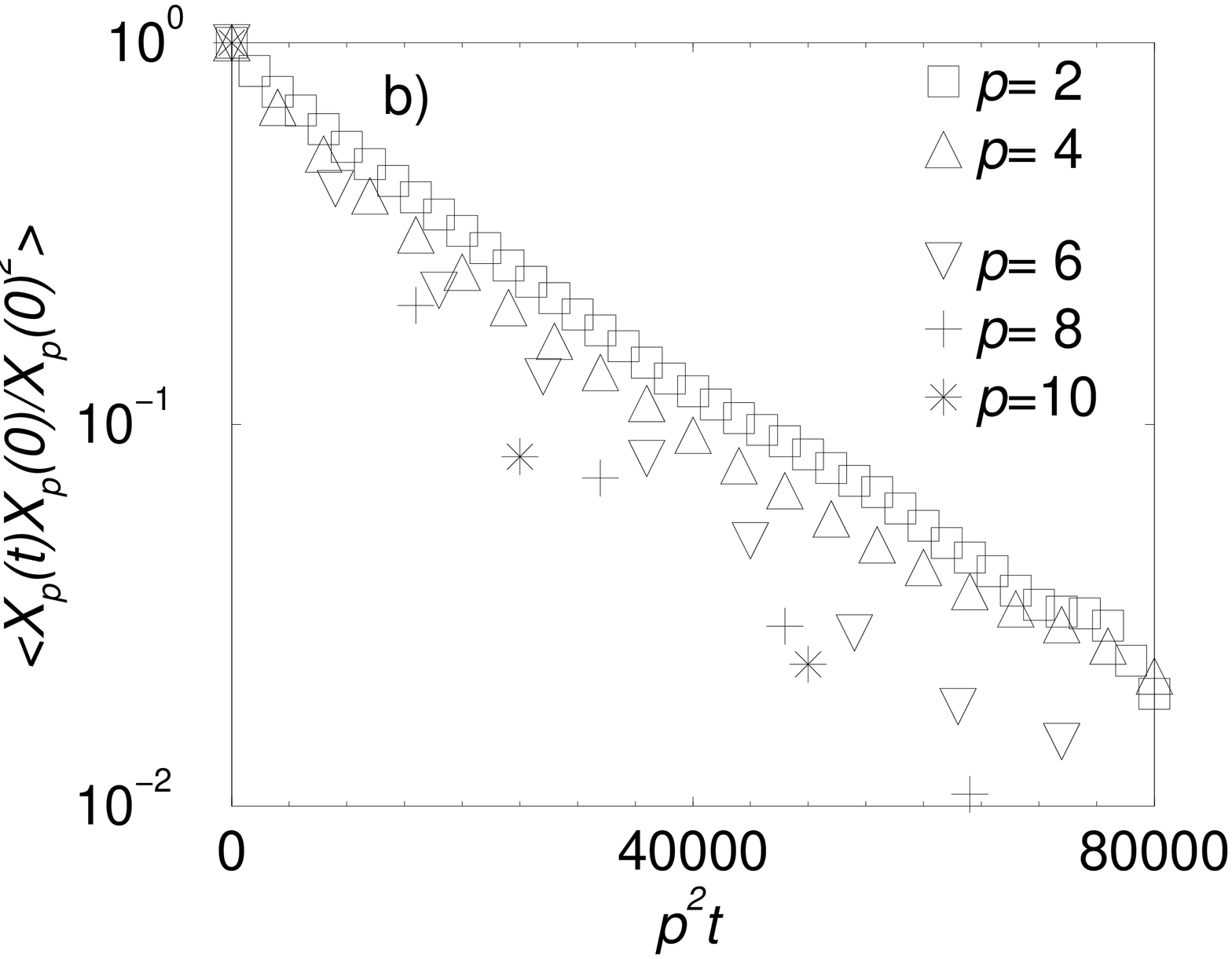}
  \includegraphics[angle=-90,width=0.5\linewidth]{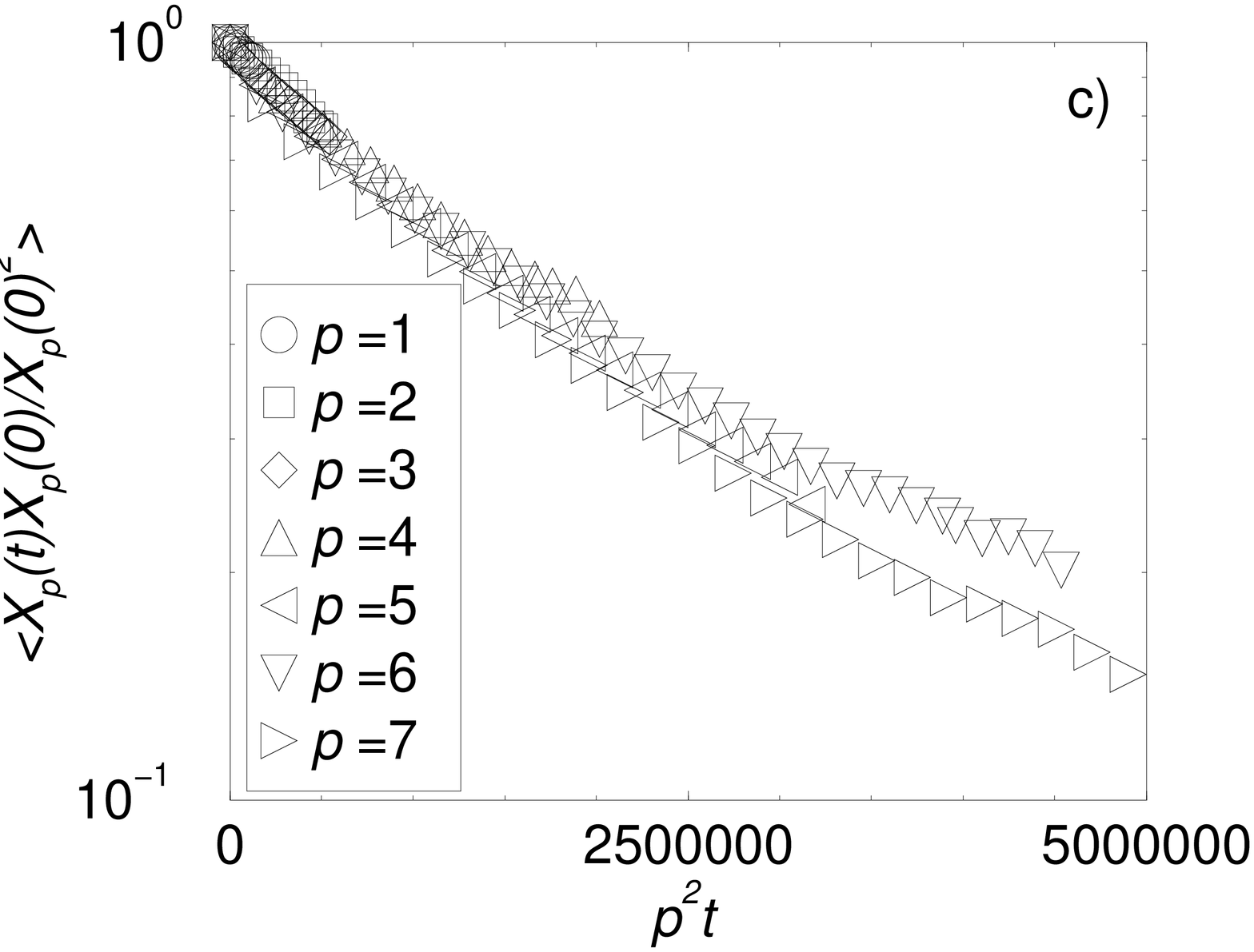}
  \includegraphics[angle=-90,width=0.5\linewidth]{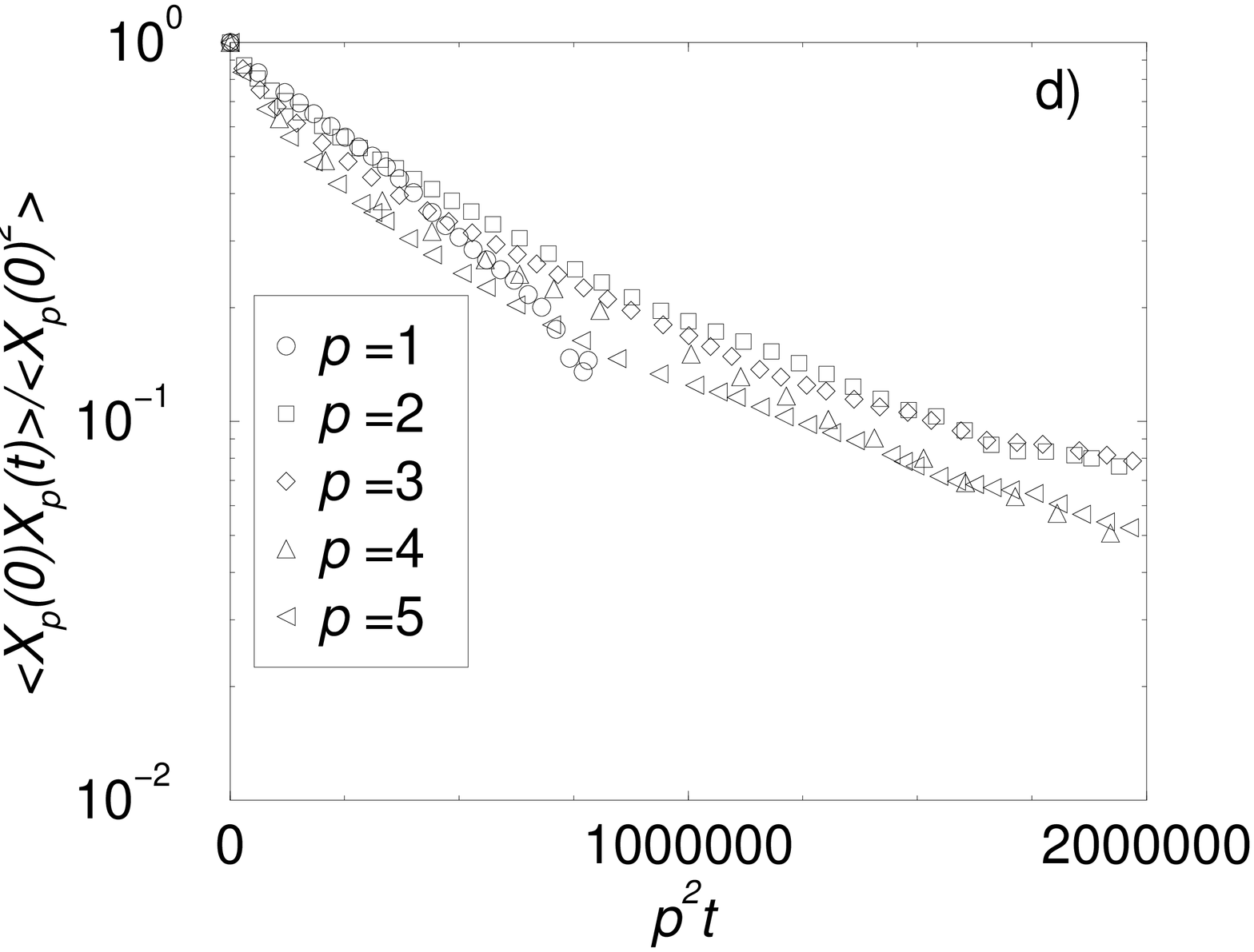}
  \caption{Normalized correlation functions of Rouse modes $X_p$ of stiff 
    and flexible chains of various
    stiffnesses (a) 1-50; b) 3-50; c) 5-200; d) 1-350). The notation means 
    $l_p-N$ with persistence length and chain monomer number.
    The Rouse model breaks down with increasing length and stiffness. } 
  \label{fig:rousemodes}
\end{figure}

For simple flexible bead-spring chains the reptation concept was
successfully validated by several
simulations~\cite{kremer90,duenweg98,puetz00}. However, NMR
experiments of polymer reorientation show that this model cannot
describe all polymers satisfactorily. For poly-(dimethylsiloxane),
PDMS, which is known to be very flexible the results of double quantum NMR
can be explained by the simple reptation model~\cite{callaghan98}
whereas for polybutadiene this simple explanation shows strong
deficiencies~\cite{graf98}. For this purpose the reorientation
behaviour of entangled melts depending on stiffness was
investigated~\cite{faller00b}. With increasing stiffness the
reorientation slows down tremendously going hand in hand with a
decreasing entanglement length and shrinking tube
diameter~\cite{faller00sa}.  The reorientation of backbone segments is
algebraic on short time scales (cf. figure~\ref{fig:reoralg}). This
algebraic dependence $C_{reor}\propto t^{-\gamma}$ is the same as
found in NMR experiments. The exponents of $\gamma=-\frac{1}{4}$ and
$\gamma=-\frac{1}{2}$ as seen in figure~\ref{fig:cmpnmr} are both
found. However, there is a {\it qualitative} change in dynamics with
chain length as the entanglement length is crossed. The exponent of
$\gamma=-\frac{1}{4}$ is not found in unentangled chains.
\begin{table}
  \begin{center}
  \begin{tabular}{lr}
    \hline
    $l_p$ & $N_e$ \\
    \hline
    1 & 32\\
    1.5 & 15 \\
    3 & 8 \\
    5 & 6 \\
    \hline
  \end{tabular}
  \caption{Entanglement monomer number $N_e$  depending on persistence length 
    $l_p$, determined
    by the change of diffusion behavior with chain length~\cite{faller00sa}. }
  \label{tab:le}
  \end{center}
\end{table}
In diffusion also the entanglement length scale is found. Chains
longer than $l_e$ diffuse clearly slower than predicted by simple
Rouse motion (Figure~\ref{fig:ne}). According to the Rouse model the
overall diffusion is expected to be $D\propto N^{-1}$. This would
correspond to a horizontal straight line in
figure~\ref{fig:ne}. Reptation leads to $D\propto N^{-2}$ which we
find for longer chains. The crossover point can be taken as a
definition for the entanglement length~\cite{faller00sa} (Table~\ref{tab:le}).
We observe that the entanglement length {\it decreases} with increasing 
persistence length.
$l_e$ and $l_p$ are neither independent nor linearly dependent on each other.
This makes it impossible to renormalize stiff chains onto the simple 
bead-spring model, as only one length scale could be scaled away. The
influence of stiffness {\it survives} on very long length scales.  The
entanglement length $l_e$ describes the anisotropy of motion of a chain due to
the temporary network of its uncrossable neighbors. A polymer chain has to
move predominantly along its backbone as transversal motion is hindered by the
neighbors leading to an effective tube (cf. Figure~\ref{fig:anisomot}). This
is measured by the correlation of directions of the chain (static
$\vec{u}(\tau)$) with the motion of the monomer in time 
(dynamic $\vec{v}(t-t_0)$). 
\begin{equation}
  C=\Big\langle\frac{1}{2}\Big[3\vec{u}(\tau)\vec{v}(t-t_0)\Big]\Big\rangle
\end{equation}
This function measures the correlation between the {\it static} direction of a
chain segment at a given point in time with the direction of its {\it dynamic}
motion in the time therafter. Thus, it shows that chain segments  move in the
beginning preferably along their contour. This is exactly what reptation is 
about.
Increasing stiffness supports this effect as the stiffness suppresses
the transversal motion even further. On short time-scales where in the
standard reptation picture isotropic motion is still possible the local
stiffness disallows this motion. Thus, the chains have to reptate from the
very beginning. This is illustrated by the mean-squared displacements of inner
monomers ($\langle x^2\rangle$ cf. Figure~\ref{fig:msdcmp}).
According to the standard reptation picture one expects the scenario
we find here only for flexible chains. On short time-scales a Rouse
motion~\cite{rouse53} is found $(\langle x^2\rangle\propto t^{0.5})$,
then the Rouse motion is constrained to the tube $(\langle
x^2\rangle\propto t^{0.25})$. After the internal degrees of freedom
are relaxed the chain as a whole moves in the tube $(\langle
x^2\rangle\propto t^{0.5})$ and finally the chains reach the free
diffusion $(\langle x^2\rangle\propto t)$. The other extreme we now
see for chains with a persistence length of five monomers. The first
two dynamical regimes are missing completely, as the Rouse motion is no more
the correct description of the polymer especially on the short scales
(see Figure~\ref{fig:rousemodes}). This regime we like to call
{\it strong reptation}. Similar results have been obtained by
Morse~\cite{morse98a,morse98b,morse98c} for chains of stronger
stiffness, who introduced the terms {\it loosely entangled} and {\it
strongly entangled} for the different systems, respectively.

Note the intersections of the different curves in
Figure~\ref{fig:msdcmp}. This shows that there are regimes where
stiffer chains diffuse even faster as the entanglement length and tube
diameter come down tremendously. This can also be seen if we look at
dynamical structure factors~\cite{faller00sa} which can measure the
tube diameter.

If the Rouse model was applicable in all sub-figures of
Figure~\ref{fig:rousemodes} all curves would fall on top of each
other as the Rouse modes
\begin{equation}
  \vec{X}_p=\frac{1}{N}\sum_{i=0}^{N-1}\cos\Big(\frac{\pi p(i+1/2)}{N}\Big)
   \vec{R}_i.
\end{equation}
with $\vec{R}_i$ the position of bead $i$ and $N$ the monomer number
would be the true eigenmodes of the system and the time in 
Figure~\ref{fig:rousemodes} is rescaled accordingly. However, we see that with
increasing stiffness and chain length
the model shows deficiencies. Because of stiffness the high modes
(local motion) start to be affected and for long chains the
entanglements hinder the low modes (large scale) motion. As the
entanglement length shrinks with stiffness the Rouse regime is {\it
eaten up} between the entangled motion on the one side and the local
stiffness on the other.
\section{Conclusions: Stiffness - A decisive characteristics}
\begin{figure}
  \[
  \includegraphics[angle=-90,width=0.5\linewidth]{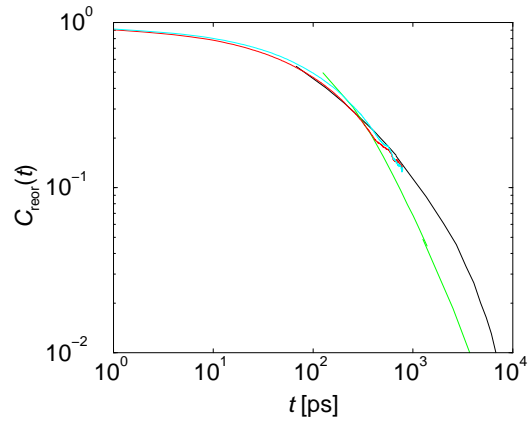}
  \]
  \caption{The reorientation of atomistic chains and simple chains 
    (both of length 10~monomers) in comparison. The red (C$_1$) and the blue 
    (C$_2$) line are atomistic vectors connecting the indicated carbons on 
    neighboring monomers. The green line is the reorientation in the simple
    model with no added stiffness. The black line corresponds to the simple 
    model with persistence length $l_p=1.5$ similar to the persistence length
    of the polyisoprene model. To suppress end-effects the terminal monomers 
    were not taken into account.}  
  \label{fig:cmp}
\end{figure}

Two different polymer models were introduced. A bead-spring model with
stiffness and a fully detailed atomistic model. Both are validated
against different experiments. The connecting link between the models
is the backbone stiffness which {\it survives} from the very local
scale to the global scale on which often only entanglements are
expected to be important. In Figure~\ref{fig:cmp} one sees the success
of the mapping. The atomistic chains at 413~K and model chains (also
of length~10) are compared (one monomer to one monomer mapping). The
mapping is accomplished by rescaling (squared) lengths with the
mean-squared end-to-end distance. Time-scales are fixed by the
center-of-mass diffusion. Then the figure shows the comparative
reorientation of local monomer-to-monomer vectors. Thus, this study
opens one possible route to polymer coarse graining: Simulate an
atomistic melt (of short chains) in full detail, there you can investigate all
the very
local observables (sub-monomer to monomer-scale) like radial
distribution functions, orientation correlations up to structure
functions and reorientation times. Also one has to determine the
persistence length of the polymer on this scale. This is then
taken as an input to the model on the next length scale, so that part of the
polymer identity is preserved and a tremendous simulation speedup is
possible at the same time. At this scale, long-time dynamic phenomena
like reptation and large scale static structure like the overall
Gaussian random walk distribution can be looked into.

However, one has to be very careful with this mapping. There are polymers for
which this route is too simple-minded, especially if the monomer is strongly
anisotropic or has special or bulky side-groups. More elaborate methods have
to be applied then~\cite{tschoep98a,meyer00}. Our method can be 
applied to dense melts of simple hydrocarbon polymers and allows in this case
a very strong speedup and a look on real large scale phenomena. With
the other methods typically an intermediate scale between the two presented
scales here is introduced as additional length scales can become important. 
Still, for dynamic issues it is not possible even on the largest scale to 
explain everything with simple chains only, at least stiffness has to be taken 
into account. 
\section*{Acknowledgments}
Many fruitful discussions with Markus Deserno, Burkhard D{\"u}nweg, Ralf
Everaers,  Andreas Heuer, Kurt Kremer,
Heiko Schmitz and Doros Theodorou gave valuable ideas to our
work. Additionally, financial support by the German Ministry of Research
(BMBF) is gratefully acknowledged.

\bibliography{standard}
\bibliographystyle{CTPS}

\end{document}